\documentclass[twocolumn,preprintnumbers,amsmath,amssymb]{revtex4}
\usepackage{graphicx,subfigure,epsfig,epstopdf}
 \usepackage{graphicx}
\usepackage{dcolumn}
\usepackage{bm}
\usepackage{amsmath,amssymb}
\input{epsf}
\input{epsfx}

\def\braket#1{\mathinner{\langle{#1}\rangle}}

\newcommand{\unit}[1]{\hspace{-1.3pt} {#1}}

\renewcommand{\baselinestretch}{1.0}
 
\begin{document}
\title{Time-resolved lasing action from single and coupled photonic crystal nanocavity array lasers emitting in the telecom-band}


\author{Dirk Englund}
\affiliation{Department of Applied Physics, Stanford University, Stanford CA 94305 }
\author{Hatice Altug}
\affiliation{Electrical and Computer Engineering Department, Boston University, Boston MA 02215}
\author{Jelena Vu\v{c}kovi\'{c}}
\affiliation{Ginzton Laboratory, Stanford University, Stanford CA 94305}

\begin{abstract}
We measure the lasing dynamics of single and coupled photonic crystal nanocavity array lasers fabricated in the indium gallium arsenide phosphide material system. Under short optical excitation, single cavity lasers produce pulses as fast as 11\unit{ps} (FWHM), while coupled cavity lasers show significantly longer lasing duration which is not explained by a simple rate equations model. A Finite Difference Time Domain simulation including carrier gain and diffusion suggests that asynchronous lasing across the nanocavity array extends the laser's pulse duration. 
\end{abstract}
\maketitle
\clearpage


The ultrasmall mode volume and high quality factor $Q$ of cavities in photonic crystals (PCs) enables controllable light-matter interaction.  This control can improve the performance of lasers by simultaneously increasing the spontaneous emission rate into the cavity mode while suppressing emission into other all other modes, resulting in a large spontaneous emission coupling efficiency $\beta$ \cite{2005.Science.Noda.SE_control,2005.PRL.Englund, 2006.PRL.Strauf,2007.NPhoton.Noda.SE,2007.OpEx.Nozaki-Baba}. Nanocavity lasers moreover enable very broad modulation bandwidth as the relaxation oscillation can be shifted beyond the cavity cutoff frequency\cite{1991.IEEE.Yamamoto}. We recently demonstrated single and coupled PC nanocavity array lasers in GaAs membranes with InGaAs quantum wells (QWs), emitting at 940-980\unit{nm} with pulse duration shorter than 4\unit{ps} (FWHM ) at room temperature\cite{2007.APL.Englund_Thz,Altug2006Nature}.  There has been particular interest in PC lasers emitting in the transparency window of standard telecommunications fiber near 1550 nm\cite{Painter99science,2000.APL.Hwang, 2001.ElectronLett.Monat}. We have recently addressed this wavelength band with PC nanocavity array lasers in the InGaAsP material system, emitting from 1530-1550\unit{nm} in quasi-continuous mode operation\cite{2005.OpEx.Altug}. To increase power output from single cavities, we also described a coupled PC nanocavity array design. In this letter, we investigate the time-domain lasing characteristics of such coupled- and single cavity PC lasers in the InGaAsP material system, to better understand their lasing dynamics and potential modulation rates. Under optical excitation with 3-ps pulses above the semiconductor's bandgap energy, we measure lasing response as fast as 11\unit{ps} for the single-cavity structures.  This is explained by a three-level rate equations model. However, the coupled-cavity array laser has a longer response time, as long as 25\unit{ps}; this is not explained by the rate equations model.  We instead analyze the coupled cavity array lasers with a finite difference time domain (FDTD) simulation incorporating a carrier gain model, which suggests that the extended lasing action results from spatially non-uniform optical pumping of nanocavities and results in asynchronous lasing action near threshold. 

\begin{figure}[htbp]
 \includegraphics[width=\linewidth]{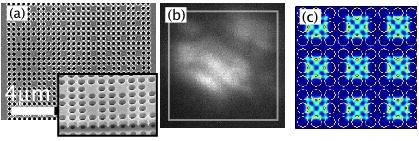} 
\caption{\footnotesize{(a) Scanning electron micrograph of coupled nanocavity photonic crystal array. The inset shows the cross-section of the PC membrane. (b) Far-field radiation pattern of coupled cavity array mode obtained at a pump power 1.4 above threshold. (c) Electric field intensity of coupled quadrupole mode}}
    \label{fig:Fig1}
\end{figure}

The structures are fabricated in a 280-nm thick In$_{0.786}$Ga$_{0.214}$As$_{0.445}$P$_{0.555}$ membrane containing four $9-$nm thick In${_0.78}$Ga$_{0.22}$As$_{0.737}$P$_{0.263}$ QWs, separated by 20-nm barriers, as described in Ref. \cite{2006.thesis.Altug}. The membrane rests on an InP substrate. Photonic crystals are created using electron beam lithography followed by a combination of wet and dry etching. We fabricated single cavities (Fig.\ref{fig:carrier_sim}(c)) and coupled cavity arrays (Fig.\ref{fig:Fig1}(a)) in square lattice PCs with periodicity $a=500$nm and hole radii ranging from $160$\unit{nm} to $230$\unit{nm}. The array contains 9x9 cavities that are spaced by two holes.  It supports a coupled quadrupole mode (Fig.\ref{fig:Fig1}(c)) that is designed to overlap with the QW gain. 

The structures were tested in a confocal microscope setup at room temperature.  The QWs were excited by pumping with a pulsed Ti:Sapph laser at an 80 MHz repetition rate with a pulse duration of 3.5 ps pulses and a center wavelength at 770 nm, above the bandgap energy of the In$_{0.786}$Ga$_{0.214}$As$_{0.445}$P$_{0.555}$ membrane.  The emission was measured using an optical spectrum analyzer and a streak camera (Hamatsu N5716-02) for time-resolved measurements.

\begin{figure}[htbp]\centering
\includegraphics[width=\linewidth]{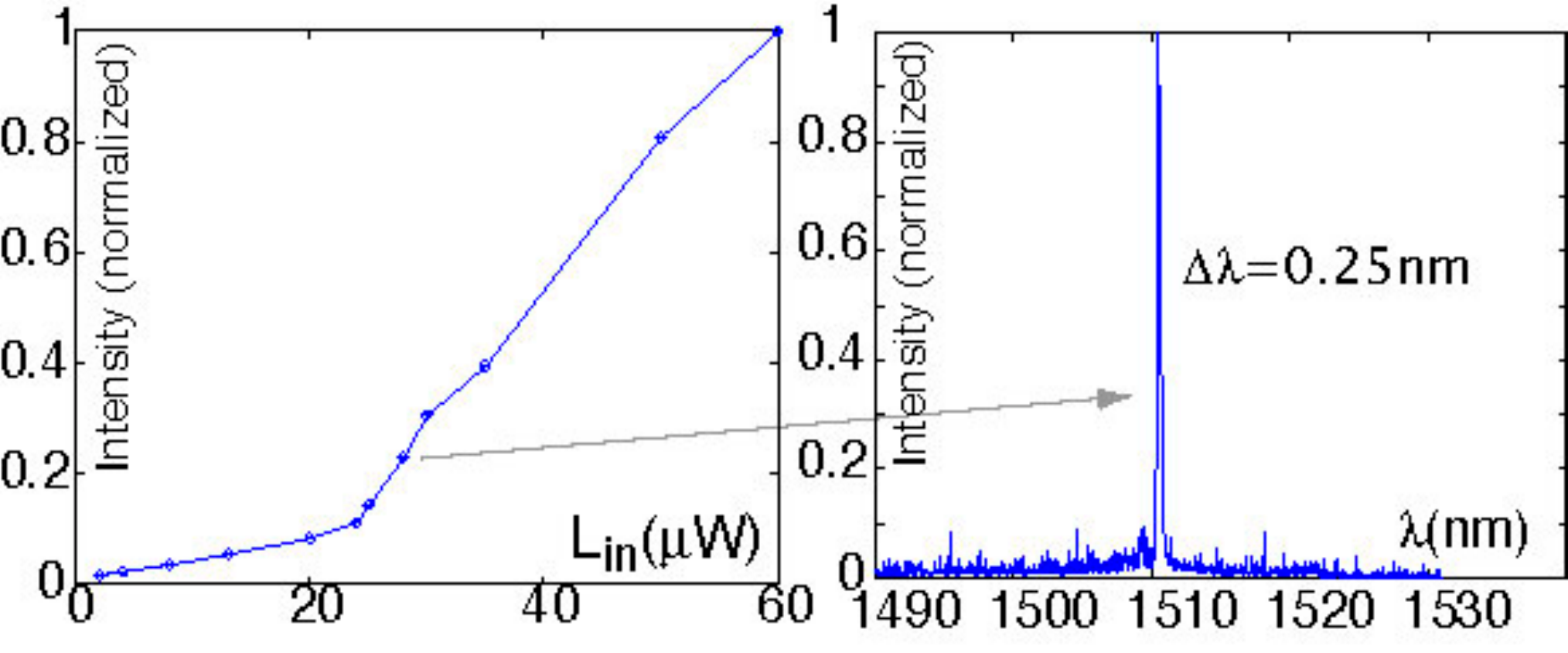} 
\caption{{\footnotesize (a) LL curve of single-cavity structure under pulsed excitation. (b) Single-cavity laser spectrum at $28\mu$W averaged pump power.  }}
    \label{fig:Fig2}
\end{figure}

\begin{figure}[htbp]\centering
\includegraphics[width=\linewidth]{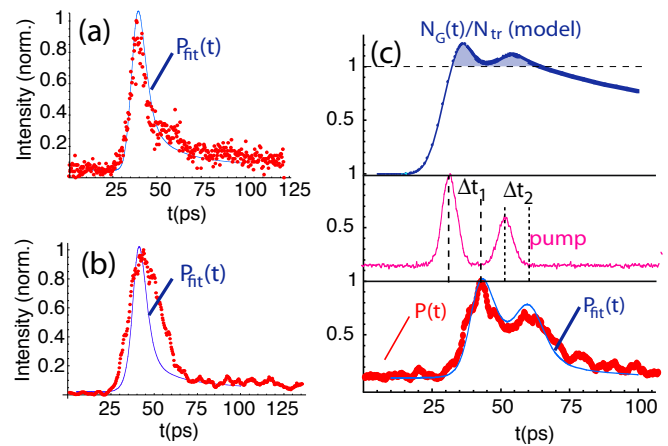}
    \renewcommand{\baselinestretch}{1.0}
\caption{{\footnotesize (a) Single-cavity lasing response (dots) and rate equations fit $P_{\mbox{fit}}$. Pump power $\braket{L_{in}}=31$\unit{$\mu$W}. All plots are normalized to the maximum intensity. (b) Coupled-cavity array lasing response ($\braket{L_{in}}=1.4\times$ threshold). The rate equations model does not adequately explain the long lasing duration. (c) Single-cavity response to excitation by two pulses: top, lasing level concentration; center, measured excitation sequence; bottom, observed intensity and rate equations fit. }}
\label{fig:Fig3}
\end{figure}

\begin{figure}[htbp]\centering
\includegraphics[width=\linewidth]{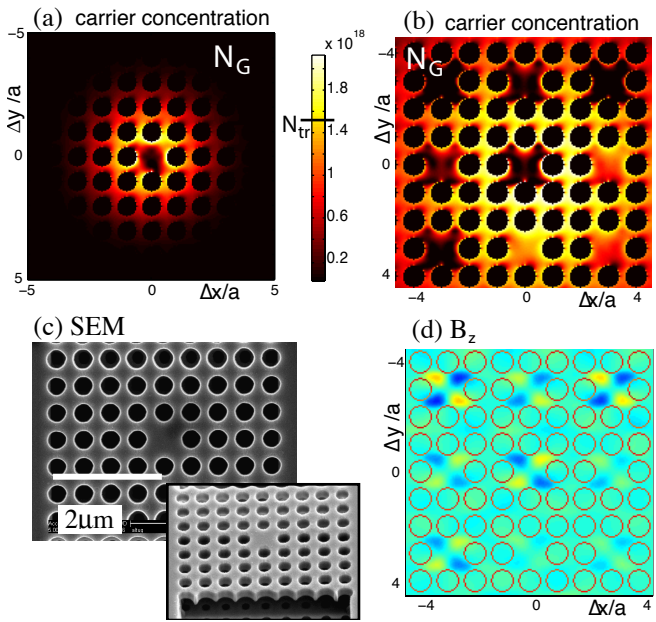}
    \renewcommand{\baselinestretch}{1.0}
\caption{{\footnotesize  (a) Lasing-level carrier concentration (1 ps after injection) showing density gradient towards lasing cavity.  Spatial hole burning results from the fast stimulated recombination during the lasing pulse. Pump power is $2\times N_{tr}$ at the center of the gaussian spot with radius $2 a$, where $a$ is the PC lattice period. (b) Carrier concentration in PC array, 1\unit{ps} after injection.  Pump energy corresponds to $1.4\times N_{tr}$. Small inhomogeneities in the pump spot (radius $6 a$) and coupled cavity mode can result in a spreading of lasing onset times, contributing to longer total pulse duration. (c) SEM of single-cavity InP laser structure. (d) Out-of-plane magnetic field of lasing mode, 1\unit{ps} after carrier injection. }}
\label{fig:carrier_sim}
\end{figure}

The lasing response of a typical single-cavity structure behavior is shown in the light-in/light out (LL) curve in Fig. \ref{fig:Fig2}(a) and indicates a threshold of $\braket{L_{in}}=22$\unit{$\mu$W} time-averaged power (corresponding to $\sim 71$mW peak power in a 3.5\unit{ps} pulse).  At an average pump power of $\braket{L_{in}}=28\mu$W, we observe a lasing mode with a FWHM of $0.25$\unit{nm} at 1511\unit{nm} (Fig.\ref{fig:Fig2}(b)). A streak-camera measurement at 30$\mu$W indicates a lasing response with FWHM $\sim 11$\unit{ps}. This is shown in Fig.\ref{fig:Fig3}(a).  We analyze the lasing dynamics using the rate equations model described in Ref.\cite{2008.LPR.Englund.Laser_review}.  Briefly, this model assumes a homogeneous distribution of carries across the spatial extent of the PC structure. The carriers are considered to be either in the ground, the pump, or in the lasing level. Using parameters derived from experiment, together with literature values for the gain and transparency carrier concentration\footnote{Gain is modeled as $g(n_G)=\Gamma g_0 v_g \ln(n_G/V_a N_{tr})$, $V_a=$active material volume, $g_0\approx 1500$/cm, group velocity $v_g=c/n_{eff},n_{eff}=2.6$,$N_{tr}\approx 1.5\cdot 10^{18}$\unit{cm$^{-3}$}(from \cite{1995Coldren}); gain confinement factor $\Gamma\approx 0.159$}, the model predicts a cavity photon number $P_{\mbox{fit}}(t)$ (convolved with the streak camera response) which is in good agreement with the experimental data in Fig.\ref{fig:Fig3}(a).

To estimate the maximum modulation rate of the single-cavity laser, we excited the structure with a series of pulses produced with an etalon setup in the excitation path. A small angle misalignment in the etalon caused consecutive pulses to walk off from the excitation path, so that only two pulses are visible in the streak-camera measurement of the pump (center panel of Fig.\ref{fig:Fig3}(c)). The laser is pumped at 1.4 times above the threshold, where it showed stable operation. The pulse separation shown is $21$\unit{ps} and represents the smallest pulse separation that resulted in two clearly distinguishable lasing response pulses (bottom panel of Fig.\ref{fig:Fig3}(c)). This pulse repetition is longer than the sub- $10$ps repetition time reported for GaAs lasers with InGaAs QWs\cite{Altug2006Nature}. The laser turn-on delay is $\tau_1=11$\unit{ps} for the first pulse and $\tau_2=9$ \unit{ps} for the second pulse. To understand the lasing dynamics, we again model the system with the rate equations model.  The result is shown in the solid curve $P_{\mbox{fit}}(t)$ and fits the data well. It is useful to consider the lasing level concentration $N_G(t)$ predicted by the model; it is plotted in the top panel of Fig.\ref{fig:Fig3}(c), normalized by the carrier transparency concentration $N_{tr}\sim 1.5\cdot 10^{18}$cm$^{-3}$. The fraction of carriers that exceeds the transparency value (shaded region) is efficiently converted to cavity photons during the lasing process, since the threshold carrier concentration roughly equals the transparency concentration; the remainder recombines primarily through nonradiative recombination at the photonic crystal hole boundaries, with a recombination time given by $\tau_{nr}=r/2S$, where $r$ and $S$ are the hole radius and the surface recombination velocity, respectively\cite{Hayes1988APL,2007.APL.Englund_pass}. This decay is also visible in the spontaneous emission tail in the bottom panel of Fig.\ref{fig:Fig3}(c). From separate lifetime measurements, we estimate $\tau_{nr}\approx 270$\unit{ps}, which gives $S\approx 3\cdot 10^{4}$cm/s.  This value is 2-3 times higher than the value of $S\sim 10^4$cm/s reported elsewhere for this material system\cite{2001.APL.Baba.surface_recomb,2005.OpEx.Altug}, probably due to the processing of the holes. The slower decay of carriers that remain after the first lasing pulse (roughly equal to $N_{tr}$ in the top panel of Fig.\ref{fig:Fig3}(c)) strongly impacts the dynamics of the second laser pulse, and would lead to eye-closing under a pseudo-random bit sequence. Therefore, return-to-zero signaling would be problematic at a modulation rate exceeding $\sim 1/\tau_{NR}$ unless the laser is pumped far above threshold where the remaining carrier concentration $N_{tr}$ becomes insignificant. In our suspended structures, thermal problems made it difficult to achieve stable lasing action above $\sim 1.6$ times the threshold power. Heat dissipation can be improved by fabricating the PC laser structures on top of low-index substrates such as sapphire or silicon oxide\cite{2000.APL.Hwang,2001.ElectronLett.Monat,2003.JAP.Monat,2007.OpEx.Raj,2006.OpEx.Bakir-Fedeli}. 

We next turn to the 9x9 photonic crystal nanocavity array. At a pump power of 2 times above threshold and below, we measured significantly longer lasing duration.  For a pump intensity of $1.4$ times above threshold, where we achieved stable operation, we measured FWHM$\approx 19$\unit{ps} (Fig.\ref{fig:Fig3}(b)). This longer response time is not adequately explained by the rate equations model, as shown in the poor match of the best fit to the data. 

We believe that the poor fit to the rate equations model arises in large part because it does not account for spatial variations in the carrier concentration across the photonic crystal device.  It was found previously that the spatial profile significantly impacts lasing dynamics, for example through spatial hole burning\cite{2004.APL.Baba}, and is important in understanding lasing threshold\cite{2007.APL.Englund_pass}.  To understand its role in the laser time response, we have implemented carrier dynamics in our finite-difference time domain model. Such nonlinear FDTD implementations have been used previously to model the dynamics in PC lasers\cite{2007.JLTech.Pernice}. Material gain is implemented in FDTD by an effective conductivity $\sigma$, as in references \cite{2004.JOptB.Kretschmann,1996.RadioSci.Hagness}. An auxiliary differential equation is used to describe the evolution of the current density $\vec{J}$.  In turn, $\vec{J}$ is related to the carrier density $N_G$ (assumed here to be equal for holes and electrons).  The set of equations obtained when $\vec{J}=\sigma \vec{E}$ is substituted into Maxwell's equations is then expressed in the time-domain and discretized as described in \cite{2007.JLTech.Pernice}.   

A simulation of optical pumping with a gaussian-shaped beam results in inhomogeneous gain and asynchronous lasing action, spreading out the total pulse duration.  This is seen from the lasing level concentration $N_G$ and cavity field in Fig. \ref{fig:carrier_sim}(b,d), recorded here 1\unit{ps} after injecting carriers at 1.4 times the transparency concentration in the center. The cavities are in different stages of the lasing cycle: in some cavities, the gain has already been used up (dark cavities in (b)), while other cavities are still at the onset of lasing (bright cavities in (b)). The corresponding cavity fields are shown in the $B_z$ component of the field in (d). The carrier concentration in (b) is blurred through carrier diffusion with a ambipolar diffusion constant of $7$cm$^{2}/$s\cite{2000.IEEE.Button.diffusion}. As a result of the asynchronous lasing action across the coupled-cavities array, the nanocavities are not phase-locked together, and the total response time is broadened. This hypothesis is supported by the CCD image in Fig.\ref{fig:Fig1}(b), which shows that the laser emission at $1.4\times$ threshold is not uniform across the structure. Experimentally it appears that at higher pump power, the pulse response becomes shorter; unfortunately, the lasing response becomes unstable when the pump power exceeds roughly 1.4 times the threshold power, so we were not able to acquire reliable data on the streak camera, probably due to heating problems. Nevertheless, this observation would support our model, as all cavities would reach lasing threshold more rapidly.    

In conclusion, we have measured time-resolved lasing action of single and coupled nanocavity lasers in the InGaAsP/InP material system and emitting near the telecommunications band. Single-cavity lasers show lasing response as fast as 11\unit{ps} (FWHM) at $1.5\times$ above threshold power, but their modulation rate appears $\sim 2\times$ slower than that of InGaAs/GaAs PC lasers, probably due to slower nonradiative recombination. The significantly longer response time of the coupled cavity array structure indicates that under excitation with short pulses, it is difficult to excite the full photonic crystal structure uniformly to achieve phase-locking across all nanocavities. This result suggests that large-signal modulation of the coupled cavity array laser requires attention to uniform injection current. 

This work was supported by the MARCO Interconnect Focus Center and the NSF. 
%



\begin{thebibliography}{10}

\bibitem{2005.Science.Noda.SE_control}
Masayuki Fujita, Shigeki Takahashi, Yoshinori Tanaka, Takashi Asano, and Susumu
  Noda.
\newblock {Simultaneous Inhibition and Redistribution of Spontaneous Light
  Emission in Photonic Crystals}.
\newblock {\em Science}, 308(5726):1296--1298, 2005.

\bibitem{2005.PRL.Englund}
D.~Englund, D.~Fattal, E.~Waks, G.~Solomon, B.~Zhang, T.~Nakaoka, Y.~Arakawa,
  Y.~Yamamoto, and J.~Vu\v{c}kovi\'{c}.
\newblock {Controlling the Spontaneous Emission Rate of Single Quantum Dots in
  a Two-Dimensional Photonic Crystal}.
\newblock {\em Phys. Rev. Lett.}, 95:013904, July 2005.

\bibitem{2006.PRL.Strauf}
S.~Strauf, K.~Hennessy, M.~T. Rakher, Y.-S. Choi, A.~Badolato, L.~C. Andreani,
  E.~L. Hu, P.~M. Petroff, and D.~Bouwmeester.
\newblock Self-tuned quantum dot gain in photonic crystal lasers.
\newblock {\em Phys. Rev. Lett.}, 96(12):127404, 2006.

\bibitem{2007.NPhoton.Noda.SE}
S.~Noda, M.~Fujita, and T.~Asano.
\newblock {Spontaneous-emission control by photonic crystals and nanocavities}.
\newblock {\em Nature Photonics}, 1:449 -- 458, 2007.

\bibitem{2007.OpEx.Nozaki-Baba}
Kengo Nozaki, Shota Kita, and Toshihiko Baba.
\newblock Room temperature continuous wave operation and controlled spontaneous
  emission in ultrasmall photonic crystal nanolaser.
\newblock {\em Opt. Express}, 15(12):7506--7514, 2007.

\bibitem{1991.IEEE.Yamamoto}
G.~Bjork and Y.~Yamamoto.
\newblock {Analysis of semiconductor microcavity lasers using rate equations}.
\newblock {\em IEEE Journal of Quantum Electonics}, 27(11):2386--96, November
  1991.

\bibitem{2007.APL.Englund_Thz}
D.~Englund, H.~Altug, I.~Fushman, and J.~Vu\v{c}kovi\'{c}.
\newblock {Efficient Terahertz Room-Temperature Photonic Crystal Nanocavity
  Laser}.
\newblock {\em Appl. Phys. Lett.}, 91:071126, July 2007.

\bibitem{Altug2006Nature}
H.~Altug, D.~Englund, and J.~Vu\v{c}kovi\'{c}.
\newblock {Ultrafast photonic crystal nanocavity laser}.
\newblock {\em Nature Physics}, 2:484--488, 2006.

\bibitem{Painter99science}
O.~Painter, R.K. Lee, A.~Scherer, A.~Yariv, J.D. O'Brien, P.D. Dapkus, and
  I.~Kim.
\newblock {Two-Dimensional Photonic Band-Gap Defect Mode Laser}.
\newblock {\em Science}, 284:1819--1821, June 1999.

\bibitem{2000.APL.Hwang}
Jeong-Ki Hwang, Han-Youl Ryu, Dae-Sung Song, Il-Young Han, Hyun-Woo Song,
  Hong-Kyu Park, Yong-Hee Lee, and Dong-Hoon Jang.
\newblock Room-temperature triangular-lattice two-dimensional photonic band gap
  lasers operating at 1.54 mu m.
\newblock {\em Appl. Phys. Lett.}, 76(21):2982--2984, 2000.

\bibitem{2001.ElectronLett.Monat}
C.~Monat, C.~Seassal, X.~Letartre, P.~Viktorovitch, P.~Regreny, M.~Gendry,
  P.~Rojo-Romeo, G.~Hollinger, E.~Jalaguier, S.~Pocas, and B.~Aspar.
\newblock \mbox{InP} 2d photonic crystal microlasers on silicon wafer: room
  temperature operation at 1.55 um.
\newblock {\em Electronics Letters}, 37(12):764--766, 7 Jun 2001.

\bibitem{2005.OpEx.Altug}
H.~Altug and J.~Vu\v{c}kovi\'{c}.
\newblock Photonic crystal nanocavity array laser.
\newblock {\em Opt. Express}, 13(22):8819--8828, 2005.

\bibitem{2006.thesis.Altug}
H.~Altug.
\newblock {\em {Physics and Applications of Photonic Crystal Nanocavities}}.
\newblock PhD thesis, Stanford University, December 2006.

\bibitem{2008.LPR.Englund.Laser_review}
D.~Englund, H.~Altug, Bryan Ellis, and J.~Vuckovic.
\newblock Ultrafast photonic crystal lasers.
\newblock {\em Laser \& Photonics Reviews}, pages 1863--8880, 2008.

\bibitem{Hayes1988APL}
K.~Tai, T.~R. Hayes, S.~L. McCall, and W.~T. Tsang.
\newblock {Optical measurement of surface recombination in \mbox{InGaAs}
  quantum well mesa structures}.
\newblock {\em Appl. Phys. Lett.}, 53(4):302--303, July 1988.

\bibitem{2007.APL.Englund_pass}
D.~Englund, H.~Altug, and J.~Vu\v{c}kovi\'{c}.
\newblock {Low-Threshold Surface-Passivated Photonic Crystal Nanocavity Laser}.
\newblock {\em Appl. Phys. Lett.}, 91:071124, July 2007.

\bibitem{2001.APL.Baba.surface_recomb}
Hiroyuki Ichikawa, Kyoji Inoshita, and Toshihiko Baba.
\newblock Reduction in surface recombination of gainasp microcolumns by ch[sub
  4] plasma irradiation.
\newblock {\em Applied Physics Letters}, 78(15):2119--2121, 2001.

\bibitem{2003.JAP.Monat}
C.~Monat, C.~Seassal, X.~Letartre, P.~Regreny, M.~Gendry, P.~Rojo Romeo,
  P.~Viktorovitch, M.~Le~Vassor d'Yerville, D.~Cassagne, J.~P. Albert,
  E.~Jalaguier, S.~Pocas, and B.~Aspar.
\newblock Two-dimensional hexagonal-shaped microcavities formed in a
  two-dimensional photonic crystal on an \mbox{InP} membrane.
\newblock {\em Journal of Applied Physics}, 93(1):23--31, 2003.

\bibitem{2007.OpEx.Raj}
G.~Vecchi, F.~Raineri, I.~Sagnes, A.~Yacomotti, P.~Monnier, T.~J. Karle, K.-H.
  Lee, R.~Braive, L.~Le Gratiet, S.~Guilet, G.~Beaudoin, A.~Taneau,
  S.~Bouchoule, A.~Levenson, and R.~Raj.
\newblock Continuous-wave operation of photonic band-edge laser near 1.55 um on
  silicon wafer.
\newblock {\em Opt. Express}, 15(12):7551--7556, 2007.

\bibitem{2006.OpEx.Bakir-Fedeli}
B.~B. Bakir, C.~Seassal, X.~Letartre, P.~Regreny, M.~Gendry, P.~Viktorovitch,
  M.~Zussy, L.~Di Cioccio, and J.-M. Fedeli.
\newblock Room-temperature \mbox{InAs}/\mbox{InP} quantum dots laser operation
  based on heterogeneous ``2.5 d'' photonic crystal.
\newblock {\em Opt. Express}, 14(20):9269--9276, 2006.

\bibitem{2004.APL.Baba}
T.~Baba, D.~Sano, K.~Nozaki, K.~Inoshita, Y.~Kuroki, and F.~Koyama.
\newblock Observation of fast spontaneous emission decay in gainasp photonic
  crystal point defect nanocavity at room temperature.
\newblock {\em Appl. Phys. Lett.}, 85(18):3989--3991, 2004.

\bibitem{2007.JLTech.Pernice}
W.~H. Pernice, F.~P. Payne, and D.~F. Gallagher.
\newblock {Numerical investigation of field enhancement by metal nano-particles
  using a hybrid FDTD-PSTD algorithm}.
\newblock {\em Opt. Express}, 15(18):11433--11443, 2007.

\bibitem{2004.JOptB.Kretschmann}
M.~Kretschmann and A.~A. Maradudin.
\newblock {Lasing action in waveguide systems and the influence of rough
  walls}.
\newblock {\em J. Opt. Soc. Amer. B, Opt. Phys.}, 21:150Ð158, January 2004.

\bibitem{1996.RadioSci.Hagness}
S.~C. Hagness, R.~M. Joseph, and A.~Taflove.
\newblock {Subpicosecond electrodynamics of distributed Bragg reflector
  microlasers: Results from finite difference time domain simulations}.
\newblock {\em Radio Sci}, 31(4):931Ð941, 1996.

\bibitem{2000.IEEE.Button.diffusion}
D.~Marshall, A.~Miller, and C.C. Button.
\newblock In-well ambipolar diffusion in room-temperature ingaasp multiple
  quantum wells.
\newblock {\em Quantum Electronics, IEEE Journal of}, 36(9):1013--1015, Sep
  2000.

\bibitem{1995Coldren}
L.~A. Coldren and S.~W. Corzine.
\newblock {\em {Diode Lasers and Photonic Integrated Circuits}}.
\newblock Wiley, New York, 1995.

\end{thebibliography}

\end{document}